\documentclass[prl,amssymb,twocolumn,superscriptaddress]{revtex4}
\usepackage{graphicx}
\usepackage{amsmath}
\usepackage{bm}

\newcommand{\fref}[1]{Fig.~\ref{#1}}

\begin{document}

\title{Creep motion of an elastic string in a random potential}

\author{Alejandro B. Kolton, Alberto Rosso, and Thierry Giamarchi}
\affiliation{Universit\'e de Gen\`eve, DPMC, 24 Quai Ernest Ansermet, CH-1211 Gen\`eve
4, Switzerland}

\begin{abstract}
We study the creep motion
of an elastic string in a two dimensional pinning landscape by
Langevin dynamics simulations. We find that the 
Velocity-Force characteristics are well described by the 
creep formula predicted from phenomenological scaling arguments.  
We analyze the creep exponent $\mu$, and the roughness 
exponent $\zeta$. Two regimes are 
identified: when the temperature is larger than the 
strength of the disorder we find $\mu \approx 1/4$ and 
$\zeta \approx 2/3$, in agreement with the
quasi-equilibrium-nucleation picture of creep 
motion; on the contrary, lowering enough the temperature, 
the values of $\mu$ and $\zeta$ increase showing a strong 
violation of the latter picture. 
\end{abstract}


\maketitle
Understanding the physical properties of disordered elastic
systems is a challenging question relevant to a host of
experimental situations. Indeed such a situation is realized in
many different systems, ranging from periodic ones, such as vortex
lattices
\cite{blatter_vortex_review,nattermann_vortex_review,giamarchi_vortex_review},
charge density waves \cite{gruner_revue_cdw}, Wigner crystals
\cite{giamarchi_wigner_review}, to interfaces, such as magnetic
\cite{lemerle_domainwall_creep,shibauchi_creep_magnetic,caysol_minibridge_domainwall}
or ferroelectric \cite{tybell_ferro_creep} domain walls, fluid
invasion in porous media \cite{wilkinson_invasion} and domain
growth \cite{kardar_growth}. The competition between disorder and
elasticity in these systems leads to unique physical properties
and in particular to glassy behavior. One particularly important
question is the response of the system to an
external force (magnetic or electric field for domain walls,
current for vortices, etc.). At zero temperature, due to disorder
the system is pinned and the velocity of the elastic structure
remains zero up to a critical force $F_c$. At finite temperature
however, the barriers to motion due to pinning can always be
passed by thermal activation an one expects thus a finite response
at finite force. Although it was initially believed that the
response was linear \cite{anderson_kim}, it was subsequently
proposed \cite{ioffe_creep,nattermann_rfield_rbond} that due to
the glassy nature of the disordered system, no linear response
would exists. The slow dynamics of the system for $F \ll F_c$, 
so-called creep, is expected to be controlled by thermally 
activated jumps of correlated regions over the pinning energy 
barriers separating different metastable states. By adding some 
strong assumptions on
this physical picture of the motion, elegant scaling arguments
were used \cite{feigelman_collective,nattermann_pinning} to infer
the small $F$ response, leading to the ``creep formula''
\begin{equation}\label{eq:vf}
 V(F) \sim \exp \biggl[ -\frac{U_c}{T} \biggl(
 \frac{F_c}{F}\biggr)^\mu \biggr]
\end{equation}
where $U_c$ is an energy scale, and $\mu$ a characteristic
exponent that can be obtained from the relation,
\begin{equation} \label{eq:valuemu}
 \mu = \frac{D-2+2\zeta}{2-\zeta}
\end{equation}
where $D$ is the dimensionality of the elastic system, and the
exponent $\zeta$ is the \emph{equilibrium} roughness exponent of
the \emph{static} system. The above formulas are indeed derived
under the assumption that the movement is so slow that static
properties can be used. Relation (\ref{eq:valuemu}) is remarkable
since it links the statics with the nonlinear transport of a
disordered elastic system.

Going beyond the simple scaling arguments or checking for such a
law has proved to be very challenging. Although sub-linear response
was clearly seen in various systems \cite{blatter_vortex_review}
with good agreement with (\ref{eq:vf}), a precise determination of
the exponent was clearly more difficult. Relation
(\ref{eq:valuemu}) has been confirmed experimentally only for
magnetic domain walls \cite{lemerle_domainwall_creep} (see also
\cite{fuchs_creep_bglass} for vortices). On the theoretical side
the phenomenological predictions of (\ref{eq:vf}-\ref{eq:valuemu})
have been derived by a functional renormalization group
calculation \cite{chauve_creep_short,chauve_creep_long}, starting
directly from the equation of motion, and valid in an $\epsilon =
4-D$ expansion. This calculation confirmed the phenomenological
hypothesis made in the scaling derivation and the validity
of (\ref{eq:valuemu}) up to the lowest order in $\epsilon$. Although 
the velocity found in the FRG
calculation was identical to the one of the scaling derivation,
important differences were also found, notably on the
characteristic sizes involved in the motion.

In spite of these results, the physical picture of creep motion is
still very phenomenological and many important questions remain
open. A systematic study of the temperature (or disorder strength)
dependence of the creep response is particularly lacking, both
from experiments and theory. Moreover, from the theoretical point
of view, the experimentally very relevant case of low dimensional
interfaces (where thermal effects are expected to be very
important), like elastic lines describing domain walls in thin
films, posses a difficult problem to tackle
analytically, since the FRG
\cite{chauve_creep_long,muller_creep_frg} can hardly be used in
$D=1$. Such studies are quite crucial given the recent
experimental results on creep in magnetic
\cite{lemerle_domainwall_creep,shibauchi_creep_magnetic,caysol_minibridge_domainwall}
and ferroelectric \cite{tybell_ferro_creep} systems. Numerical
simulations are a valuable alternative theoretical tool to address
this open issue. In this respect, creep simulations of elastic
strings ($D=1$) have been done in the past
\cite{kaper_creep_simulation,chen_marchetti}, but given the
limited range of velocities available, they were neither systematic nor
conclusive about the validity of (\ref{eq:vf}-\ref{eq:valuemu}).

In this work we used a Langevin dynamics method to study the Velocity-Force 
(V-F) characteristics and the dynamic roughness $\zeta$ of an elastic
string in a random potential. The range of velocities we can
explore allows a precise check of the creep law. Although we find
that the creep law does describe well the data, we also find that
the equilibrium hypothesis for $\zeta$ is not verified. This leads
at low temperatures or strong disorder to creep and roughness
exponents that become larger than the predicted values $\mu=1/4$,
$\zeta=2/3$ respectively.

We study the creep motion of an elastic string 
in two dimensions driven through a random potential. 
The string is described by a single valued function $u(z,t)$, 
which measures its transverse displacement $u$ from the 
$z$ axis at a given time $t$. We therefore exclude overhangs and 
pinched-off loops that eventually could be produced 
in the motion of domain walls.
Assuming a linear short-range elasticity and 
a purely relaxational dynamics 
the phenomenological Langevin equation describing 
the motion (per unit length) is given by,
\begin{eqnarray}
\gamma \partial_t u(z,t) &=& 
c \partial_z^2 u(z,t) + F_p(u,z) + F + \eta(z,t)
\label{eqmotion}
\end{eqnarray} 
where $\gamma$ is the friction coefficient, $c$ is the elastic constant,  
$F$ is the driving force, and 
$F_p(u,z)=-\partial_u U(u,z)$ is the pinning 
force derived from the disordered pinning potential $U(u,z)$. 
The stochastic force $\eta(z,t)$ ensures a proper thermal 
equilibration and satisfies 
$\langle \eta(z,t) \rangle = 0$,  
$\langle \eta(z,t) \eta(z',t') \rangle = 2\gamma T \delta(z-z')
\delta(t-t')$, with $\langle ... \rangle$ denoting thermal average. 
The sample to sample fluctuations of the random potential are given by,
\begin{eqnarray}
\overline{ [U(u,z) -  U(u',z')]^2} &=& -2 \delta(z-z') R(u-u') 
\label{corrpinning}
\end{eqnarray} 
where the over-line denotes average over disorder realizations. In 
this work we consider a random-bond type of disorder, characterized 
by a short-ranged correlator $R(u)$, of range $r_f$ and strength 
$R(0)$. A physical realization of this kind of disorder  
is for instance the random anisotropy for magnetic 
domain walls \cite{lemerle_domainwall_creep}.

To perform numerical simulations of equation (\ref{eqmotion}) 
we discretize the string along the $z$ direction, 
$z\rightarrow j=0,\ldots,L-1$, keeping $u_j(t)$ as a 
continuous variable. A second order stochastic Runge-Kutta 
method \cite{helfand_runge_kutta,greenside_runge_kutta,honeycutt_runge_kutta} is used to integrate the resulting 
equations. To model a continuous random potential 
satisfying  (\ref{corrpinning}), we generate, for each $j$, 
a cubic spline $U(u_j,j)$ passing through $M$ regularly 
spaced uncorrelated Gaussian random points, 
with zero mean and variance $R(0)^2$ \cite{rosso_depinning_simulation}. 
Moreover the random potential satisfies periodic boundary 
conditions, 
\begin{eqnarray}
U(u_j+M,j)=U(u_j,i+L)=U(u_j,j)
\end{eqnarray} 
this defines a finite sample of size $(L,M)$.  

We are interested in the V-F characteristics and the dynamic 
roughness exponent $\zeta$ in the creep regime, for different 
values of $T$ and disorder strength $R(0)$. 
The average center of mass velocity $V$ is defined as,  
\begin{eqnarray}
V(F) = \overline{\biggl\langle \frac{1}{L}\sum_{j=0}^{L-1}\frac{d}{dt}u_j(t) \biggr\rangle}. 
\label{V}
\end{eqnarray} 
The roughness exponent $\zeta$ is obtained from the average structure factor, defined as,  
\begin{eqnarray}
S(q) = \overline{\biggl\langle \biggl|\frac{1}{L}\sum_{j=0}^{L-1} u_j(t) e^{-iqj}\biggr|^2  \biggr\rangle}
\label{S}
\end{eqnarray} 
where $q=2\pi n/L$, with $n=1,\ldots,L-1$. From dimensional analysis
we know that for small $q$, $S(q) \sim q^{-(1+2\zeta)}$ \cite{barabasi_book}. 
Fitting our numerical data with this function we extract $\zeta$. 

We simulate systems of sizes $L=64,128,\ldots,1024$ and $M=2L$, 
with $c=r_f=\gamma=1$. In this paper 
we show the results for $L=512$, where finite size effects are negligible. 
We take $R(0)=0.12, 0.30$ and  
temperatures ranging from $T=0.8 R(0)$ to $T=3.5 R(0)$. 
We start each simulation with a flat initial configuration ($u=0$) at the force 
$F=F_c/10$, and then decrease slowly $F$ in steps of $\Delta F = 0.01 F_c$ up to 
$F/F_c \approx 0.01$. $F_c$ is calculated to high precision for each 
disorder realization using a fast-convergent algorithm \cite{rosso_depinning_simulation}. 
The properties of interest have to be calculated  
when the stationary state is achieved. In practice we let the string 
complete two turns around the system for the initial force  
and one turn for the following forces. After this equilibration 
we estimate numerically (\ref{V}) and (\ref{S}) 
approximating the average over disorder and thermal realizations 
by a single time average over one turn.

\begin{figure}
 \centerline{\includegraphics[width=8.5cm]{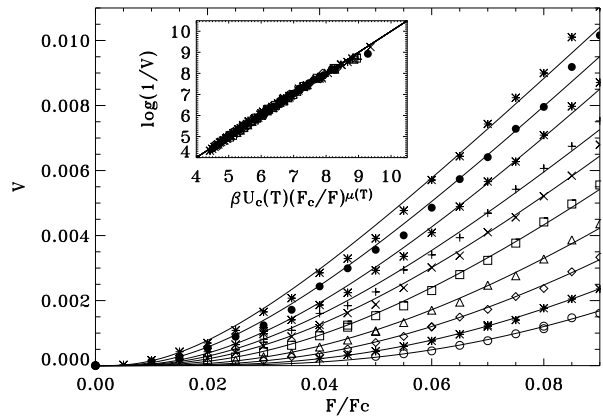}}
 \caption{V-F characteristics for $R(0)=0.30$. Curves 
 correspond to $T=0.24 ,0.26,..., 0.42$ from bottom 
 ($\circ$ symbols) to top ($*$ symbols). Solid lines 
 are fits of the creep law (\ref{eq:vf}) with $U_c$ and 
 $\mu$ as fitting parameters. Contrarily to the naive 
 creep prediction, the optimal
 fit parameter $\mu$ is temperature dependent.
 The inset shows $\log(1/V)$ vs $\beta
 U_c(F_c/F)^{\mu}$ for all $T
 $, using their respective
 fitting parameters $\mu(T)$ and $U_c(T)$. }
 \label{fig:vf}
\end{figure}

Typical V-F characteristics obtained in the simulations 
are shown in \fref{fig:vf}. 
In the whole range of temperature and pinning strength analyzed we
find that the V-F curve can be well
fitted by the creep formula (\ref{eq:vf}) with $U_c$ and $\mu$ as
fitting parameters. We thus confirm the predicted stretched
exponential behavior, being the range of
velocities in our simulations sufficient to rule out other 
proposed forms \cite{anderson_kim,vinokur_marchetti}. 
However, contrarily to the naive creep relation (\ref{eq:valuemu}) 
we find that both fitting parameters, and not only $U_c$ as 
predicted in Refs. \cite{nattermann_creep_domainwall,muller_creep_frg}, 
can depend on temperature. 

\begin{figure}
 \centerline{\includegraphics[width=8.5cm]{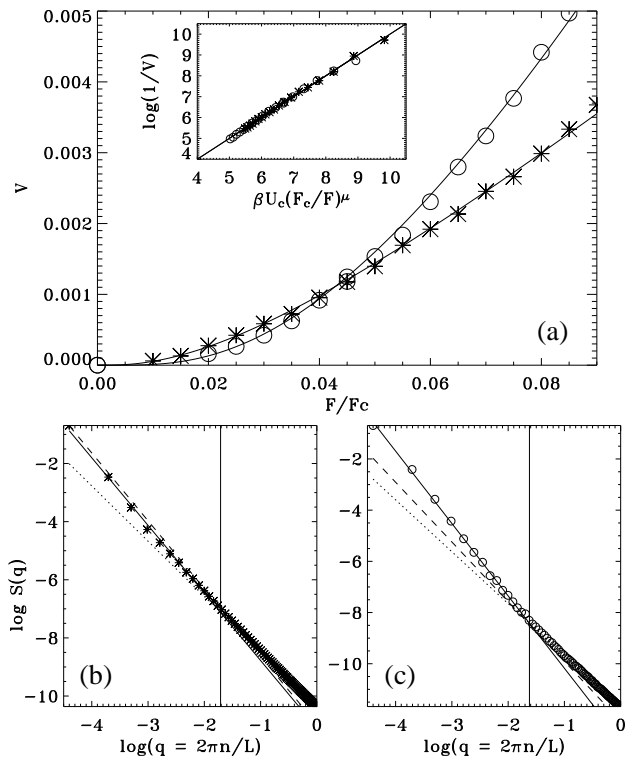}}
 \caption{(a) V-F characteristics: (i) 
 $R(0) = 0.12$, $T=0.24$ ($*$ symbols);
 (ii) $R(0) = 0.30$, $T=0.30$ ($\circ$ symbols). Solid lines are 
 the fitting curves using (\ref{eq:vf}). The inset shows
 $\log(1/V)$ vs $\beta U_c(F_c/F)^{\mu}$ using the fitting
 parameters $\mu = 0.26$, $\beta U_c = 3.1$ for (i), and 
 $\mu = 0.36$, $\beta U_c = 2.2$ for (ii).
 (b) (resp. (c)) Structure factor $S(q)$ at $F/F_c=0.02$, for (i)
 (resp. (ii)). Solid lines are fitting curves for small $q$s. We extract
 $\zeta \approx 0.67$ for (b) and $\zeta \approx 0.89$ for
 (c). Dashed (dotted) lines correspond to the reference value
 $\zeta=2/3$ ($\zeta_T=1/2$). The vertical lines indicate the 
 approximate location of the crossover from the thermal to 
 the random manifold scaling.}
 \label{twoVF}
\end{figure}

Analysis of various values of disorder and
temperature show essentially two different regimes of creep
motion. To investigate further these regimes we show 
in \fref{twoVF}(a) the V-F characteristics for 
two values of disorder and temperature 
representative of each regime.
For the small disorder case we get the
exponent $\mu = 0.26 \pm 0.01$ which is compatible with the predicted 
theoretical value $\mu = 1/4$, obtained from (\ref{eq:valuemu}) using the
equilibrium roughening exponent $\zeta_{eq} = 2/3$ 
\cite{huse_exponent_line,kardar_exponent_line}. 
The situation is quite different for the strong disorder case, 
where although the fit with the creep formula (\ref{eq:vf}) is 
still excellent, the value of the exponent $\mu \approx 0.36$ is 
now clearly in excess with respect to the predicted 
theoretical value.

To understand in more detail the nature of the 
two regimes we calculate the roughness exponent $\zeta$   
using the structure factor (\ref{S}). 
Quite generally, one can predict that the short distance behavior 
of an elastic string is dominated by thermal fluctuations 
($\zeta_T=1/2$). On the other 
hand, because of the finite velocity, the quenched disorder acts 
effectively as a thermal noise at the largest length scale. 
Thus, in this case, the expected exponent is also $\zeta_V=1/2$ 
\cite{nattermann_stepanow_depinning}.  
Finally, at intermediate length scales, the physics is 
determined by the competition between disorder and 
elasticity. In particular, in our simulations we 
verified that the Larkin length \cite{larkin_ovchinnikov_pinning} 
is negligible. Therefore, a random manifold scaling, 
characterized by a non trivial roughness exponent, takes 
place. In \fref{twoVF}(b) and (c) we show the structure 
factor for the two cases analyzed in \fref{twoVF}(a). As 
predicted, we get $\zeta \sim \zeta_{T}=1/2$ for
large $q$. At a certain scale we observe a crossover 
between the thermal and the random manifold 
scaling. The location of this crossover decreases 
as $T$ ($R(0)$) is increased (decreased). We can also 
observe that the second velocity-controlled crossover is not 
achieved in our finite-size simulation due to the 
very slow dynamics. Interestingly, for the small disorder 
case, the random manifold scaling gives 
$\zeta \approx 0.67$, in excellent agreement
with the equilibrium value $\zeta_{eq}=2/3$, while a 
much higher roughness exponent $\zeta \approx 0.89$ is 
found for the strong disorder case. These results are consistent 
with the previous ones for the creep exponent $\mu$. 
\begin{figure}
 \centerline{\includegraphics[width=8.5cm]{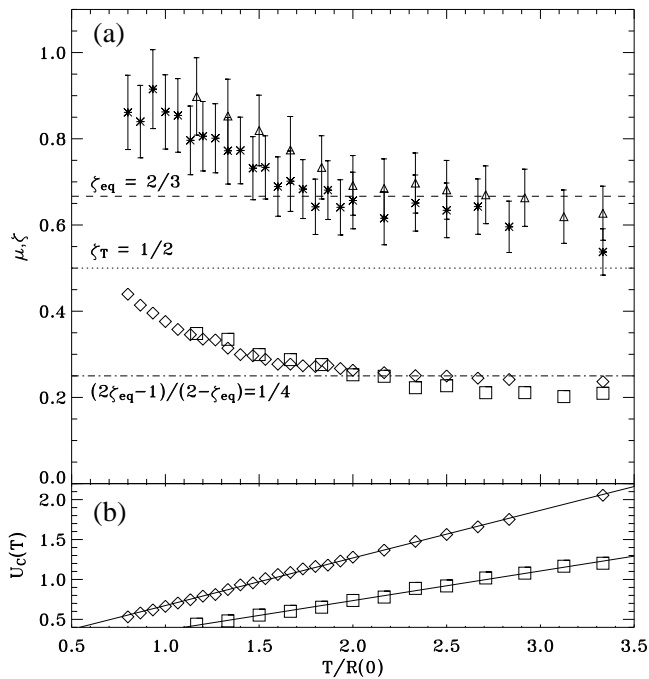}}
 \caption{(a) Roughness exponent, $\zeta(T)$, and
 creep exponent, $\mu(T)$, vs $T$. $\zeta(T)$ ($*$ symbols),
 $\mu(T)$ ($\diamond$ symbols) correspond to $R(0)=0.30$ and 
 $\zeta(T)$ ($\triangle$ symbols), $\mu(T)$ ($\square$ symbols) correspond 
 to $R(0)=0.12$. The dashed line gives
 the equilibrium roughness exponent $\zeta_{eq}=2/3$, and the dotted 
 line the purely thermal roughness $\zeta_T=1/2$. The dashed-dotted line
 corresponds to the creep exponent predicted from 
 phenomenological scaling arguments. (b) Effective energy barriers 
 $U_c(T)$ vs $T$, for $R(0)=0.30$
 ($\diamond$ symbols) and $R(0)=0.12$ ($\square$ symbols).}
 \label{fig:sum}
\end{figure}
This conclusion holds for the whole range of temperature and 
disorder strength analyzed, as we can see in \fref{fig:sum}(a). 
We find that the relevant parameter to define the two 
regimes is $T/R(0)$. It would be interesting to determine 
whether such a type of scaling has a
theoretical justification. Moreover, we notice that although the 
values of $\zeta$ and $\mu$ depart from the equilibrium values, the relation
(\ref{eq:valuemu}) seems still to hold, within the error bars for
the two exponents. This is highly 
non-trivial since equation (\ref{eq:valuemu}) is derived from a calculation
of the barriers in an equilibrium situation. 

We discuss finally the temperature dependence of the 
barriers $U_c(T)$, shown in \fref{fig:sum}(b). 
We remark that the observed linear temperature dependence is
of course peculiar to the one dimensional wall, where 
thermal fluctuations lead to unbounded displacements,
contrarily to what happens in higher dimensions 
\cite{nattermann_creep_domainwall,muller_creep_frg}.

In conclusion we have find two regimes of creep motion.
The first one occurs when the temperature is larger than 
the strength of the disorder, giving $\mu \sim 1/4$ and 
$\zeta \sim 2/3$ as predicted by assuming a quasi-equilibrium 
nucleation picture of the creep motion. This implies that 
the domain wall has time to re-equilibrate between hops, 
being the underlying assumption behind (\ref{eq:valuemu}) 
essentially satisfied. 
The second regime occurs for temperatures smaller than the 
strength of the disorder, and is characterized by anomalously large 
values of both exponents. This clearly shows that in this regime the
domain wall stays out of equilibrium, and that the naive creep
hypothesis does not apply. Note that the measured roughness
exponent is intermediate between the equilibrium value and the
depinning value $\zeta_{dep} \approx 1.2$. The fact that
the thermal nucleation which is the limiting process in the creep
velocity, is in fact followed by depinning like avalanches was
noted in the FRG study of the creep \cite{chauve_creep_long}.
Whether such avalanches and the time it would take them to relax
to equilibrium is at the root of the observed increase of the
exponent, is clearly an interesting but quite complicated
open question.

We acknowledge D. Dom\'{\i}nguez for illuminating discussions 
and C. Bolech for his careful reading of the manuscript. This work 
was supported in part by the Swiss
National Fund under Division II.


\end{document}